\documentclass[
 hf,
]{ceurart}
 \usepackage{float}

\begin{document}

\copyrightyear{2025}
\copyrightclause{Copyright for this paper by its authors.\\
  Use permitted under Creative Commons License Attribution 4.0
  International (CC BY 4.0).}

%\conference{add later}

\title{Chatting with Papers: A Hybrid Approach Using LLMs and Knowledge Graphs}


\author[1]{Vyacheslav Tykhonov}[%
orcid=0000-0001-9447-9830,
email=vyacheslav.tykhonov@dans.knaw.nl,
url=https://https://pure.knaw.nl/portal/en/persons/vyacheslav-tykhonov,
]
\address[1]{Data Archiving and Networked Services, Royal Netherlands Academy of Arts and Sciences (DANS-KNAW), 2593 HW The Hague, Anna van Saksenlaan 51, The Netherlands}

\author[2]{Han Yang}[%
orcid=0009-0000-0913-824X,
email=Han.Yang@gesis.org,
url=https://www.gesis.org/institut/ueber-uns/mitarbeitendenverzeichnis/person/Han.Yang,
]
\address[2]{GESIS -- Leibniz-Institute for the Social Sciences, Cologne, Germany,}

\author[2]{Philipp Mayr}[%
orcid=0000-0002-6656-1658,
email=philipp.mayr@gesis.org,
url=https://philippmayr.github.io/,
]

\author[1]{Jetze Touber}[%
orcid=0000-0003-4532-1273,
email=jetze.touber@dans.knaw.nl,
url=https://pure.knaw.nl/portal/en/persons/jj-touber,
]
%\address[1]{affiliation 2}

\author[1]{Andrea Scharnhorst}[%
orcid = 0000-0001-8879-8798, email = andrea.scharnhorst@dans.knaw.nl, url = https://pure.knaw.nl/portal/en/peoples/andrea-scharnhorst,]
%\address[2]{affiliation 2}

\begin{abstract}
This demo paper reports on a new workflow \textit{GhostWriter} that combines the use of Large Language Models and Knowledge Graphs (semantic artifacts) to support navigation through collections. Situated in the research area of Retrieval Augmented Generation, this specific workflow represents the creation of local and adaptable chatbots. Based on the tool-suite \textit{EverythingData} at the backend, \textit{GhostWriter} provides an interface that enables querying and ``chatting'' with a collection. Applied iteratively, the workflow supports the information needs of researchers when interacting with a collection of papers, whether it be to gain an overview, to learn more about a specific concept and its context, and helps the researcher ultimately to refine their research question in a controlled way. We demonstrate t he workflow for a collection of articles from the \textit{method data analysis} journal published by GESIS -- Leibniz-Institute for the Social Sciences. We also point to further application areas.
\end{abstract}


\conference{}

\maketitle

\section{Introduction}

Who would not love to have the possibility to navigate through a collection of documents by asking all kinds of questions to it? Traditionally, knowledge is communicated by language in a more or less codified way. On the less codified, natural language side, we find the development and application of Generative AI, which are primarily based on Large Language Models (LLMs) and mimic chatting with an ``expert" \cite{fan2024survey}.  At the other end of the spectrum of how knowledge is communicated, we find Knowledge Organization Systems (KOS) \cite{mazzocchi2018knowledge} (and more specifically, semantic artifacts \cite{corcho}). 
To be queried, the latter require some prior understanding of the structure in which the knowledge is codified. Think here about how one would have used a systematic catalog in a library in which books are indexed and ordered according to classification systems. Metaphorically, one could describe those two ends of the spectrum of how to engage with knowledge as ``asking an expert" versus ``asking a librarian".

The motivation for this explorative research is multifold. First, it starts with being aware of the ambiguity of natural language. There is power in this ambiguity: It serves serendipity and enables the emergence of associations across contexts. But this ambiguity might also lead to possible misunderstandings. 
So, to compensate for the ambiguity of natural language, one might look in the direction of more defined, codified terminology as expressed in KOS. But terms from KOS also carry a burden. By their nature, they are defined in specific contexts and usually do not easily travel across knowledge domains. 
The experimental workflow and demo we present here aspire to combine the best of both ways to communicate knowledge when it comes to information retrieval. This demo paper is scientifically located in the area of Retrieval Augmented Generation (see \ref{sec 2 Related work}). An additional motivation for developing this workflow lies in the fact that many generative AI solutions are provided by large tech companies. They are trained on specific materials, the origin of which is not always known. Additionally, these new generations of AI are based on stochastic methods, and so the results cannot be reproduced in a deterministic way. The so-called 'hallucinations', assertions that misrepresent facts and context, are one known effect of these stochastic models \cite{llm_lies}.  Despite these features, the power of Large Laguage Models (LLMs) is increasingly unfolded in all kind of research. The goal of the development of a fully functional prototype as reported in this demo paper was also to explore how data streams can be effectively connected to LLMs on a smaller scale, ensuring that relevant data segments are delivered to the model at the right time. Knowledge graphs appear to be a promising approach for sharing structured information across different models, as they include provenance and data origin, which are essential for trust and context, and combined with LLM's can enhance information retrieval.
With this demo paper, we discuss how LLM's and industry provided solutions can become embedded in workflows which can be controlled locally, so that AI is applied in a way following Open Science principles, and being reproducible and ethical. 

%how knowledge is communicated and precisely located at the scope of our collection. 

%Today, promoted by large language models (LLMs) \cite{GPT3} and vector databases, the retrieval augmented generation (RAG) \cite{RAG} is a popular paradigm to realize such a digital expert. 

Our paper starts with a short overview of some related works in section \ref{sec 2 Related work}. We introduce the use case in section \ref{sec 3 use case}. The description of the \textit{GhostWriter} workflow and some demonstrations of its abilities form the core of this paper in section \ref{sec 4 GhostWriter demo}. The paper closes with conclusions and an outlook into future research topics in section \ref{sec 5 Conclusion and Outlook}.

\section{Related work}
\label{sec 2 Related work}

Information Retrieval (IR) is the task of finding specific information from a large collection of documents \cite{schutze2008introduction} given a query. This is not only an important step in the scientific endeavor to answer a specific research question, but also, more generally, a crucial way to reduce information overload and cognitive overload \cite{informationoverload}. With the help of LLMs, augmented search generation (RAG) \cite{RAG} solves this task in a modern approach, in which documents, and particular paragraphs from them, are stored in the vector database, together with their semantic embeddings. 
Once a query is proposed, it will be converted into the embedding space and matched with the embedding vectors of the paragraphs stored in the vector database by calculating the cosine similarity. The matched paragraphs are consequently summarized by the LLM, and the query is then answered with natural language.

% GraphRAG

With the RAG method, a query could be answered appropriately with less likelihood of triggering the hallucination of the LLMs, as long as the paragraphs can be matched precisely. But there are situations in which the standard RAG performs less well. For instance, when questions require a comprehensive understanding of the documents on a more global scale, and especially in cases when the concepts which are vital to answer a query are located far apart in the documents. As an improvement, the GraphRAG \cite{GraphRAG} targets the same kind of query and constructs the paragraphs in a graph structure, where the entities and relationships are extracted and are used as the vertices and edges of a graph, together with a text description generated by the LLM. This graph is then partitioned hierarchically into subgraphs, called the graph communities. To answer a query, the graph communities are summarized by the LLM. The LLM toolbox contains scores which indicate whether this summarization answers the query. To generate the final answer, the summaries (of the graph communities) are ranked to the LLM-based score and concatenated until the token limit is reached. 
The GraphRAG method has been proven to performs well on tasks that require the information to be located in different parts of documents. However, the graph might be sparse when the entities and relationships are not explicitly explained or paraphrased, and hence the connections will break. In contrast, the workflow of this demo paper while based on the RAG paradigm, uses ontology and knowledge graphs to eventually define graph communities. This way, the workflow emphasizes the generality between entities and relationships in order to improve the matching accuracy in a wider range within the documents.

%  Other works
For the purpose of science studies, the new generation of LLMs sparks new ideas to further improve information retrieval and large corpus analysis, like Named Entity Recognition \cite{DBLP:journals/corr/abs-2408-13501}. Information retrieval is full of approaches to organize personal information spaces \cite{DBLP:conf/ecir/2024-1}. If it comes to exploring the past and present of science dynamics, a toolbox is available for bibliometric analysis of formal scholarly communication \cite{van2014visualizing, DBLP:journals/scientometrics/MayrS15}. 
Platforms have also been tested to collectively build, annotate, and analyze collections of scientific papers, targeting their metadata and full text \cite{chavalarias2021draw}. 
Some recent representations in industry include the Ai2 PaperFinder from Allen Ai\footnote{\url{https://paperfinder.allen.ai/chat}}, which assists users with an interactive workflow and clarifies the user's goal step-by-step, as well as the ScienceDirect AI tool from Elsevier\footnote{\url{https://elsevier.shorthandstories.com/sciencedirect-ai/}}, which emphasizes the transparency of the retrieved papers. 
In short, in Information Retrieval in general and in quantitative studies of the science, we find more and more explorations with tooling based in generative AI approaches.
In contrast to these efforts, our workflow is designed to be applied to an arbitrary collection of documents. It provides the user with a web interface but the majority of the tooling appears at the backend where a couple of new techniques (all connected via API services) is bundled together to enhance the provided information by data and metadata and to give feedback to the user in the form of natural language-based statements.

\section{Use case - journal \textit{mda}}
\label{sec 3 use case}
% [Han Yang] mda journal, a bit on history, size of collection, current interface

As a use case, we present a collection of social science articles from the journal \textit{mda - methods, data and analysis}. The journal itself was founded in 2007 and is published by GESIS - Leibniz Institute for the Social Sciences. The journal started originally with a mixture of German and English language articles ``on all questions important to quantitative methods in the social sciences, with a special emphasis on survey methodology"\footnote{\url{https://mda.gesis.org/}} to now be fully published in English as a Diamond Open Access journal. 
From this journal, we created a test collection of 100 random articles in PDF format and as demonstrated in the next section \ref{sec 4 GhostWriter demo}, we applied the \textit{GhostWriter} workflow on this collection\footnote{\url{https://gesis.now.museum/?page=about} Beta version, full text close reading approach implemented, workflow and interface still in development.}.

\section{GhostWriter demo}
\label{sec 4 GhostWriter demo}

\subsection{System Design}

The workflow has in essence two parts: a data extraction and enrichment pipeline, called \textit{Everything Data}, leveraging a knowledge graph and a human-computer-interface (hci) called \textit{GhostWriter} \cite{tykhonov2024next}. It makes use of a recent metadata format standard (called Croissant for Machine Learning) for datasets used for machine learning, and techniques to contextualise and semantically enrich information \cite{benjelloun2024croissant}.
If we talk about a workflow or pipeline here, one has to be aware that this is a rather simplifying label. For many of the current generative AI experiments, we find that their machinery consists of a 'bag of various tools' combined based on the tacit knowledge of engineers and researchers.\footnote{Personal communication with Arno Simons at the Workshop ``Large Language Models for the History, Philosophy and Sociology of Science'' Berlin, April 2-4 \url{https://www.tu.berlin/en/hps-mod-sci/workshop-llms-for-hpss}.} 

\begin{figure}[h]
    \centering
    \includegraphics[width=0.7\linewidth]{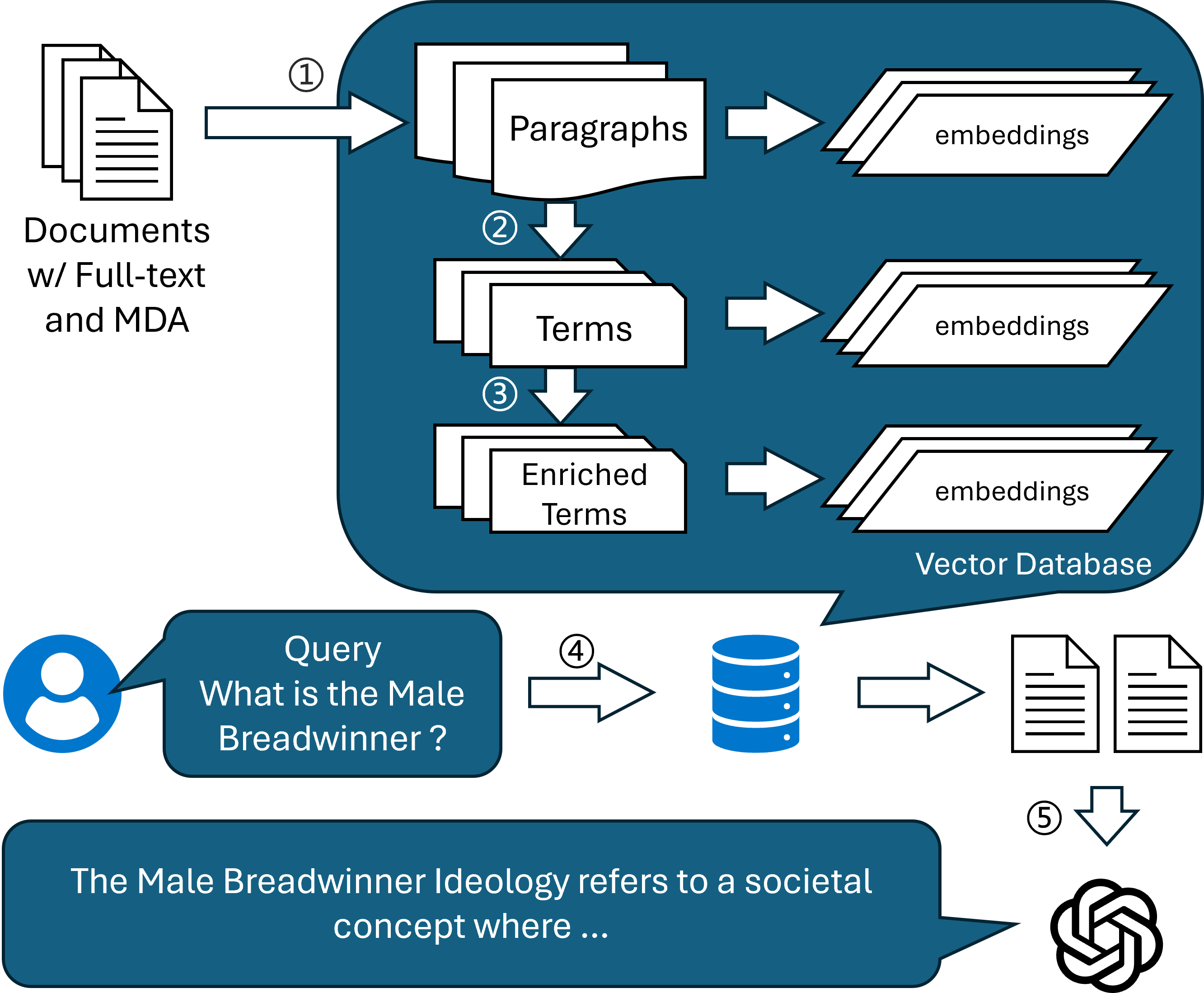}
    \caption{Schematic Workflow of \textit{EverythingData} and \textit{Ghostwriter}}
    \label{fig:workflow}
\end{figure}

For this demo paper, we decided to present elements of the workflow schematically in Figure \ref{fig:workflow}. In the upper part, we find a description of \textit{EverythingData}\footnote{https://github.com/Dans-labs/everythingdata} part of the workflow, while the lower part represents the \textit{GhostWriter} interface. \textit{EverythingData} is a generic way to extract and enrich information from documents of any kind. After documents are split and paragraphs are extracted (1, in Figure \ref{fig:workflow}), terms from these paragraphs are extracted (2) and those terms are further enriched (3). For each of these steps, embeddings are created, using a vector database in which those operations take place. As a result, we find terms which are already selected, weighted, extended according to the way the embeddings are executed, with the result that terms become contextualized further. 
In the lower part of Figure \ref{fig:workflow}, we schematically summarize the \textit{GhostWriter} interface. It starts with a query (already taken from our use case) for which information is retrieved (4). Eventually, the human-computer interface presents a natural language answer and pointers to possible source documents (5). This last step can be repeated, thus enabling an iterative way of further specifying the original query (chatting with the local collection).
The data enrichment in \textit{EverythingData} is enabled by the LLM-based term extraction and the knowledge graph \& ontology-based term enrichment, while the HCI interface is achieved by the LLMs, which enables natural language. One interesting feature of this approach is that multilinguality is naturally part of it, due to the coupling to knowledge graphs which connect various language spaces (such as wikidata). 

%https://gesis.now.museum/

\subsection{A detailed chat with \textit{method-data-analysis} papers}

\begin{figure}[h]
    \centering
    \includegraphics[width=0.9\linewidth]{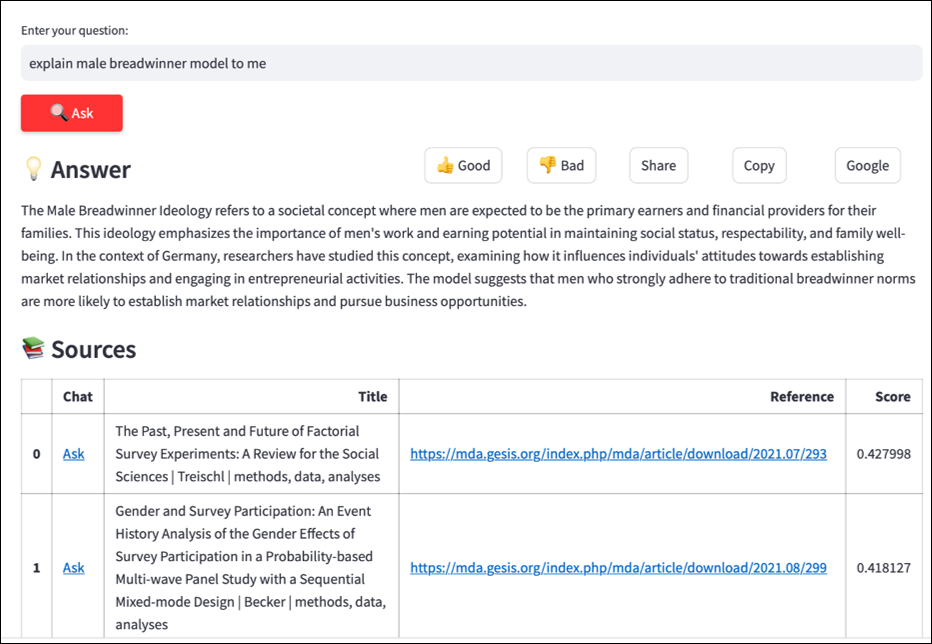}
    \caption{Question on ``male breadwinner model"}
    \label{fig:mda_1}
\end{figure}

To showcase our workflow, we created a collection with 100 articles from the \textit{mda} journal\footnote{Ghostwriter on MDA papers, betaversion, see \url{https://gesis.now.museum}, select collect collection mda}. We then proposed questions to the interface and inspected the results. For this demo paper, we choose to display the question: ``\texttt{explain male breadwinner model to me}". The first response is shown in a snapshot from the website (see Figure \ref{fig:mda_1}). One can see that the question is explained in detail, and related articles are listed as well, together with a score of confidence (calculated inside the workflow). This indicates that this workflow can answer questions in an academic language and support further investigation into the collection. 

\begin{figure}[h]
    \centering
    \includegraphics[width=1.0\linewidth]{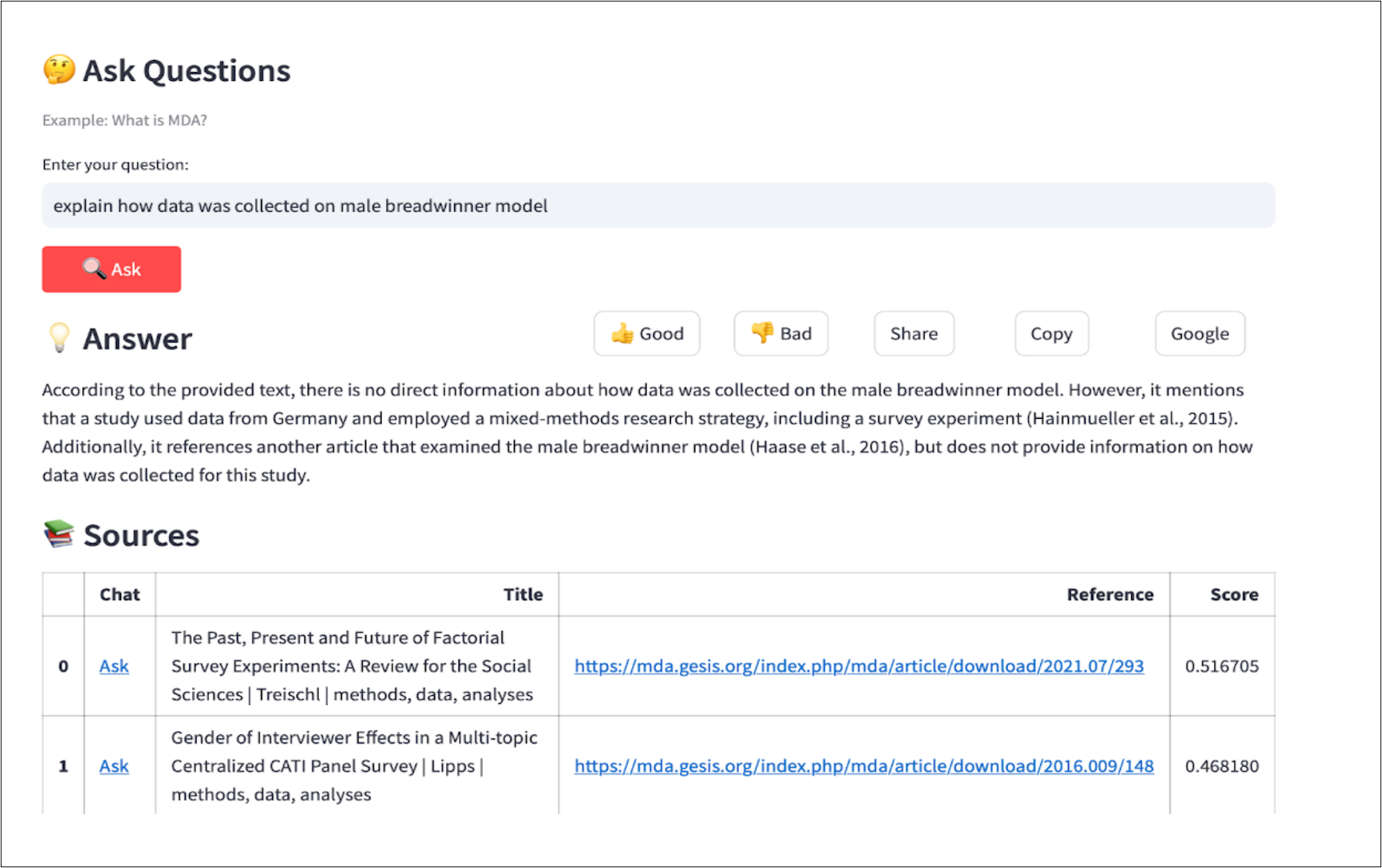}
    \caption{Finetuning the question on the ``male breadwinner model"}
    \label{fig:mda_2}
\end{figure}

It is interesting to note that the listed articles do not necessarily need to contain the exact query term to be found relevant. The user is invited to further close-reading to explore the relevance of the listed hits. 

But what is even more interesting is the fact that the question can be refined, as shown in Figure \ref{fig:mda_2}. In the provided source documents, there is no direct information about how data was collected in connection with the male breadwinner model. Still, if we refine the question in this direction, we get answers. For instance, one of the listed articles mentions that a study used data from Germany and employed a mixed-methods research strategy, including a survey experiment \cite{hainmueller2015validating}. Additionally, it references another article that examined the male breadwinner model  \cite{haase2016male}, but does not provide information on how data was collected for this study. Obviously, the associations to other terms created in the \textit{EverythingData} part of the workflow lead to suggestions to other sources. Knowledge-based enrichment of terms with their multilingual properties can increase the number of sources that can be queried and improve recall without compromising precision.
In general, this demonstration shows that a way to chat with papers can be created that goes far beyond any query term or related query term matching. To evaluate such workflows in a proper scientific manner is not an easy task and requires many well-defined user tests and evaluations, but at least we can demonstrate a working technology which can be used for the multilingual sources.

\section{Conclusion and Outlook}
\label{sec 5 Conclusion and Outlook}

This demo paper bases on experiences gained in a series of project-based explorations concerning information processes in the social sciences and humanities, and more specifically in the area cultural heritage and their information infrastructural needs, such as the European project MuseIT\footnote{\url{https://www.muse-it.eu}} \cite{johansson2025knowledge} and the Dutch research infrastructure project ODISSEI\footnote{\url{https://odissei-data.nl}}. 

As we demonstrated in Section \ref{sec 4 GhostWriter demo}, the workflow belongs to the RAG approach and relies on entity extraction, where terms are annotated with semantic meaning by mapping them into knowledge graphs built from controlled vocabularies. These ways to connect vector spaces (LLM's) to a knowledge graph (structured knowledge, KOS) open possibilities to enhance the semantic meaning for information retrieval tools. 

The challenge with this approach - which in essence is combining semantically related pieces of text - is achieving a coherent interpretation of the context. It often fails when contradictions or differing points of view are present in the source collection. Additionally, semantic similarity does not guarantee logical consistency, and models may incorrectly link unrelated content due to superficial lexical overlap. There's also a risk of the wrong understanding of the context, where nuanced or domain-specific meanings are lost.

As a possible future solution, the current ranking algorithm should be revised to first cluster and rank document fragments at the document level. This means grouping semantically similar fragments by their source document before evaluating their relevance. By doing this, we avoid mixing contradictory content from different documents and preserve contextual coherence. Once ranked, selected clusters can be passed to the LLM, along with explicit metadata about provenance such as paper titles, authors, publication dates, and source repositories. This enhances the reliability of the generated responses by making the origins of information traceable and allowing the model to distinguish between perspectives from different sources.
Linking entities to even more knowledge graph representations (such as wikidata) brings the possibility of multilingual search as one immediate benefit of such enrichment.  

As demonstrated in the case of chatting with (some) \textit{mda} papers, any query input starts a kind of distant reading machinery which returns with an answer (using natural language) together with pointers to the sources from the original collection supporting this provisional answer. This way, it mimics a close-reading level and still enables the user to continue her/his close-reading journey. 

The \textit{GhostWriter} workflow works for any ``collection". You can create a ``semantic index" and a local chatbot for collections from webpages/websites, RSS feeds, but also from trusted data repositories such as Dataverse instances. (see DataChat\footnote{https://github.com/gdcc/datachat}). In essence, you enable local chatting with papers -- this way pointing a human to specific texts which one can then engage with by way of close-reading. You confine the ‘search space’ by placing your question (and the collection) into a particular area in the networked space of scientific knowledge. We also demonstrated that you gain information beyond what is explicitly expressed in the text and annotated in the metadata (by creating associations on the level of natural language and on the level of Knowledge Organisation Systems). It is like you start chatting with the experts or invisible colleges behind the papers. 
But, there are also many preconditions to make such a workflow work. It starts with API's from which collections can be harvested, and API's which provide KOS services (for the risks of KOS services see e.g. \cite{zeng2021knowledge}). Well-curated metadata and digitized full text are another precondition. Implementation of the Croissant Standard \cite{benjelloun2024croissant} as a new standard for datasets for machine-learning is another precondition. 
As part of our future work, we have developed the Model Context Protocol (MCP)\footnote{https://mcp.dataverse.org} and initiated the practice of depositing metadata for papers separately in the Dataverse data repository. This setup allows MCP to be used as an automated factual query source. The approach appears more reliable than other approaches, as it retrieves up-to-date metadata directly from data sources and enables verification of the information provided. It’s also currently being deployed as “Ask Dataverse” services and tested by Harvard IQSS on https://ask.dataverse.org. 
So in short, while this technology offers great opportunities, it is far from being a 'ready-from-the-shelf-implementable' software. It is still very much in development.

Having said this, one large application field is collections of documents from the Cultural Heritage - for instance, the many collections that Galleries, Libraries, Archives and Musea hold \cite{8709832} contain. Developing skills to develop your workflows and to tailor them to your own needs is a very important way to empower Global Open Research Commons \cite{jones_2023}, a way to execute Open Science, even more needed today.

\section*{Acknowledgments}

Parts of this paper have been funded by projects such as ODISSEI (see \url{https://odissei-data.nl}, SSHOC.nl (NWO fund), FAIRImpact (EC Grant DOI 10.3030/101057344), MuseIT (EC Grant DOI
10.3030/101061441) and NFDI4DataScience. We thank Yves Rozenholc (University Paris Descartes) and the now.museum for their collaboration and support.
Han Yang received funding from the Deutsche Forschungsgemeinschaft (DFG) under grant number: MA 3964/15-3 (SocioHub project). Han Yang and Philipp Mayr received additional funding from the European Union under the Horizon Europe grant OMINO -- Overcoming Multilevel INformation Overload\footnote{\url{https://ominoproject.eu/}} under grant number 101086321 \cite{informationoverload}.

%%
%% Define the bibliography file to be used
\bibliography{sample-ceur}

\begin{thebibliography}{21}
\expandafter\ifx\csname natexlab\endcsname\relax\def\natexlab#1{#1}\fi
\providecommand{\url}[1]{\texttt{#1}}
\providecommand{\href}[2]{#2}
\providecommand{\path}[1]{#1}
\providecommand{\DOIprefix}{doi:}
\providecommand{\ArXivprefix}{arXiv:}
\providecommand{\URLprefix}{URL: }
\providecommand{\Pubmedprefix}{pmid:}
\providecommand{\doi}[1]{\href{http://dx.doi.org/#1}{\path{#1}}}
\providecommand{\Pubmed}[1]{\href{pmid:#1}{\path{#1}}}
\providecommand{\bibinfo}[2]{#2}
\ifx\xfnm\relax \def\xfnm[#1]{\unskip,\space#1}\fi
%Type = Inproceedings
\bibitem[{Fan et~al.(2024)Fan, Ding, Ning, Wang, Li, Yin, Chua, and Li}]{fan2024survey}
\bibinfo{author}{W.~Fan}, \bibinfo{author}{Y.~Ding}, \bibinfo{author}{L.~Ning}, \bibinfo{author}{S.~Wang}, \bibinfo{author}{H.~Li}, \bibinfo{author}{D.~Yin}, \bibinfo{author}{T.~Chua}, \bibinfo{author}{Q.~Li},
\newblock \bibinfo{title}{A survey on {RAG} meeting llms: Towards retrieval-augmented large language models},
\newblock in: \bibinfo{editor}{R.~Baeza{-}Yates}, \bibinfo{editor}{F.~Bonchi} (Eds.), \bibinfo{booktitle}{Proceedings of the 30th {ACM} {SIGKDD} Conference on Knowledge Discovery and Data Mining, {KDD} 2024, Barcelona, Spain, August 25-29, 2024}, \bibinfo{publisher}{{ACM}}, \bibinfo{year}{2024}, pp. \bibinfo{pages}{6491--6501}. \DOIprefix\doi{10.1145/3637528.3671470}.
%Type = Article
\bibitem[{Mazzocchi(2018)}]{mazzocchi2018knowledge}
\bibinfo{author}{F.~Mazzocchi},
\newblock \bibinfo{title}{Knowledge organization system (kos): an introductory critical account},
\newblock \bibinfo{journal}{KO Knowledge Organization} \bibinfo{volume}{45} (\bibinfo{year}{2018}) \bibinfo{pages}{54--78}. \URLprefix \url{https://www.isko.org/cyclo/kos}.
%Type = Article
\bibitem[{Corcho et~al.(2023)Corcho, Ekaputra, Heibi, Jonquet, Micsik, Peroni, and Storti}]{corcho}
\bibinfo{author}{{\'{O}}.~Corcho}, \bibinfo{author}{F.~J. Ekaputra}, \bibinfo{author}{I.~Heibi}, \bibinfo{author}{C.~Jonquet}, \bibinfo{author}{A.~Micsik}, \bibinfo{author}{S.~Peroni}, \bibinfo{author}{E.~Storti},
\newblock \bibinfo{title}{A maturity model for catalogues of semantic artefacts},
\newblock \bibinfo{journal}{CoRR}  (\bibinfo{year}{2023}). \DOIprefix\doi{10.48550/ARXIV.2305.06746}.
%Type = Article
\bibitem[{Yao et~al.(2023)Yao, Ning, Liu, Ning, and Yuan}]{llm_lies}
\bibinfo{author}{J.~Yao}, \bibinfo{author}{K.~Ning}, \bibinfo{author}{Z.~Liu}, \bibinfo{author}{M.~Ning}, \bibinfo{author}{L.~Yuan},
\newblock \bibinfo{title}{{LLM} lies: Hallucinations are not bugs, but features as adversarial examples},
\newblock \bibinfo{journal}{CoRR}  (\bibinfo{year}{2023}). \DOIprefix\doi{10.48550/ARXIV.2310.01469}.
%Type = Book
\bibitem[{Sch{\"u}tze et~al.(2008)Sch{\"u}tze, Manning, and Raghavan}]{schutze2008introduction}
\bibinfo{author}{H.~Sch{\"u}tze}, \bibinfo{author}{C.~D. Manning}, \bibinfo{author}{P.~Raghavan}, \bibinfo{title}{Introduction to information retrieval}, volume~\bibinfo{volume}{39}, \bibinfo{publisher}{Cambridge University Press Cambridge}, \bibinfo{year}{2008}.
%Type = Article
\bibitem[{Ho{\l}yst et~al.(2024)Ho{\l}yst, Mayr, Thelwall, Frommholz, Havlin, Sela, Kenett, Helic, Rehar, Ma{\v{c}}ek et~al.}]{informationoverload}
\bibinfo{author}{J.~A. Ho{\l}yst}, \bibinfo{author}{P.~Mayr}, \bibinfo{author}{M.~Thelwall}, \bibinfo{author}{I.~Frommholz}, \bibinfo{author}{S.~Havlin}, \bibinfo{author}{A.~Sela}, \bibinfo{author}{Y.~N. Kenett}, \bibinfo{author}{D.~Helic}, \bibinfo{author}{A.~Rehar}, \bibinfo{author}{S.~R. Ma{\v{c}}ek}, et~al.,
\newblock \bibinfo{title}{Protect our environment from information overload},
\newblock \bibinfo{journal}{Nature Human Behaviour} \bibinfo{volume}{8} (\bibinfo{year}{2024}) \bibinfo{pages}{402--403}. \DOIprefix\doi{10.1038/s41562-024-01833-8}.
%Type = Inproceedings
\bibitem[{Lewis et~al.(2020)Lewis, Perez, Piktus, Petroni, Karpukhin, Goyal, K{\"{u}}ttler, Lewis, Yih, Rockt{\"{a}}schel, Riedel, and Kiela}]{RAG}
\bibinfo{author}{P.~Lewis}, \bibinfo{author}{E.~Perez}, \bibinfo{author}{A.~Piktus}, \bibinfo{author}{F.~Petroni}, \bibinfo{author}{V.~Karpukhin}, \bibinfo{author}{N.~Goyal}, \bibinfo{author}{H.~K{\"{u}}ttler}, \bibinfo{author}{M.~Lewis}, \bibinfo{author}{W.~Yih}, \bibinfo{author}{T.~Rockt{\"{a}}schel}, \bibinfo{author}{S.~Riedel}, \bibinfo{author}{D.~Kiela},
\newblock \bibinfo{title}{Retrieval-augmented generation for knowledge-intensive {NLP} tasks},
\newblock in: \bibinfo{editor}{H.~Larochelle}, \bibinfo{editor}{M.~Ranzato}, \bibinfo{editor}{R.~Hadsell}, \bibinfo{editor}{M.~Balcan}, \bibinfo{editor}{H.~Lin} (Eds.), \bibinfo{booktitle}{Advances in Neural Information Processing Systems 33: Annual Conference on Neural Information Processing Systems 2020, NeurIPS 2020, December 6-12, 2020, virtual}, \bibinfo{year}{2020}. \URLprefix \url{https://proceedings.neurips.cc/paper/2020/hash/6b493230205f780e1bc26945df7481e5-Abstract.html}.
%Type = Article
\bibitem[{Edge et~al.(2024)Edge, Trinh, Cheng, Bradley, Chao, Mody, Truitt, Metropolitansky, Ness, and Larson}]{GraphRAG}
\bibinfo{author}{D.~Edge}, \bibinfo{author}{H.~Trinh}, \bibinfo{author}{N.~Cheng}, \bibinfo{author}{J.~Bradley}, \bibinfo{author}{A.~Chao}, \bibinfo{author}{A.~Mody}, \bibinfo{author}{S.~Truitt}, \bibinfo{author}{D.~Metropolitansky}, \bibinfo{author}{R.~O. Ness}, \bibinfo{author}{J.~Larson},
\newblock \bibinfo{title}{From local to global: A graph rag approach to query-focused summarization},
\newblock \bibinfo{journal}{arXiv preprint arXiv:2404.16130}  (\bibinfo{year}{2024}).
%Type = Article
\bibitem[{Tong et~al.(2024)Tong, Smirnova, Upadhyaya, Yu, Culbert, Sun, Otto, and Mayr}]{DBLP:journals/corr/abs-2408-13501}
\bibinfo{author}{X.~Tong}, \bibinfo{author}{N.~Smirnova}, \bibinfo{author}{S.~Upadhyaya}, \bibinfo{author}{R.~Yu}, \bibinfo{author}{J.~H. Culbert}, \bibinfo{author}{C.~Sun}, \bibinfo{author}{W.~Otto}, \bibinfo{author}{P.~Mayr},
\newblock \bibinfo{title}{Utilizing large language models for named entity recognition in traditional chinese medicine against {COVID-19} literature: Comparative study},
\newblock \bibinfo{journal}{CoRR}  (\bibinfo{year}{2024}). \DOIprefix\doi{10.48550/ARXIV.2408.13501}.
%Type = Proceedings
\bibitem[{Goharian et~al.(2024)Goharian, Tonellotto, He, Lipani, McDonald, Macdonald, and Ounis}]{DBLP:conf/ecir/2024-1}
\bibinfo{editor}{N.~Goharian}, \bibinfo{editor}{N.~Tonellotto}, \bibinfo{editor}{Y.~He}, \bibinfo{editor}{A.~Lipani}, \bibinfo{editor}{G.~McDonald}, \bibinfo{editor}{C.~Macdonald}, \bibinfo{editor}{I.~Ounis} (Eds.), \bibinfo{title}{Advances in Information Retrieval - 46th European Conference on Information Retrieval, {ECIR} 2024, Glasgow, UK, March 24-28, 2024, Proceedings, Part {I}}, volume \bibinfo{volume}{14608} of \textit{\bibinfo{series}{Lecture Notes in Computer Science}}, \bibinfo{publisher}{Springer}, \bibinfo{year}{2024}. \DOIprefix\doi{10.1007/978-3-031-56027-9}.
%Type = Incollection
\bibitem[{Van~Eck and Waltman(2014)}]{van2014visualizing}
\bibinfo{author}{N.~J. Van~Eck}, \bibinfo{author}{L.~Waltman},
\newblock \bibinfo{title}{Visualizing bibliometric networks},
\newblock in: \bibinfo{booktitle}{Measuring scholarly impact: Methods and practice}, \bibinfo{publisher}{Springer}, \bibinfo{year}{2014}, pp. \bibinfo{pages}{285--320}.
%Type = Article
\bibitem[{Mayr and Scharnhorst(2015)}]{DBLP:journals/scientometrics/MayrS15}
\bibinfo{author}{P.~Mayr}, \bibinfo{author}{A.~Scharnhorst},
\newblock \bibinfo{title}{Scientometrics and {Information} {Retrieval} - weak-links revitalized},
\newblock \bibinfo{journal}{Scientometrics} \bibinfo{volume}{102} (\bibinfo{year}{2015}) \bibinfo{pages}{2193--2199}. \DOIprefix\doi{10.1007/s11192-014-1484-3}.
%Type = Article
\bibitem[{Chavalarias et~al.(2021)Chavalarias, Lobb{\'e}, and Delano{\"e}}]{chavalarias2021draw}
\bibinfo{author}{D.~Chavalarias}, \bibinfo{author}{Q.~Lobb{\'e}}, \bibinfo{author}{A.~Delano{\"e}},
\newblock \bibinfo{title}{Draw me science: Multi-level and multi-scale reconstruction of knowledge dynamics with phylomemies},
\newblock \bibinfo{journal}{Scientometrics}  (\bibinfo{year}{2021}) \bibinfo{pages}{1--31}.
%Type = Article
\bibitem[{Tykhonov et~al.(2024)Tykhonov, van Rijsselberg, and Endarto}]{tykhonov2024next}
\bibinfo{author}{V.~Tykhonov}, \bibinfo{author}{F.~van Rijsselberg}, \bibinfo{author}{E.~Endarto},
\newblock \bibinfo{title}{The next generation of data management with artificial intelligence}  (\bibinfo{year}{2024}). \DOIprefix\doi{10.5281/zenodo.14507120}.
%Type = Article
\bibitem[{Benjelloun et~al.(2024)Benjelloun, Simperl, Marcenac, Ruyssen, Conforti, Kuchnik, van~der Velde, Oala, Vogler, Akthar, Jain, and Tykhonov}]{benjelloun2024croissant}
\bibinfo{author}{O.~Benjelloun}, \bibinfo{author}{E.~Simperl}, \bibinfo{author}{P.~Marcenac}, \bibinfo{author}{P.~Ruyssen}, \bibinfo{author}{C.~Conforti}, \bibinfo{author}{M.~Kuchnik}, \bibinfo{author}{J.~van~der Velde}, \bibinfo{author}{L.~Oala}, \bibinfo{author}{S.~Vogler}, \bibinfo{author}{M.~Akthar}, \bibinfo{author}{N.~Jain}, \bibinfo{author}{S.~Tykhonov},
\newblock \bibinfo{title}{Croissant format specification}  (\bibinfo{year}{2024}). \URLprefix \url{https://mlcommons.github.io/croissant/docs/croissant-spec.html}.
%Type = Article
\bibitem[{Hainmueller et~al.(2015)Hainmueller, Hangartner, and Yamamoto}]{hainmueller2015validating}
\bibinfo{author}{J.~Hainmueller}, \bibinfo{author}{D.~Hangartner}, \bibinfo{author}{T.~Yamamoto},
\newblock \bibinfo{title}{Validating vignette and conjoint survey experiments against real-world behavior},
\newblock \bibinfo{journal}{Proceedings of the National Academy of Sciences} \bibinfo{volume}{112} (\bibinfo{year}{2015}) \bibinfo{pages}{2395--2400}.
%Type = Article
\bibitem[{Haase et~al.(2016)Haase, Becker, Nill, Shultz, and Gentry}]{haase2016male}
\bibinfo{author}{M.~Haase}, \bibinfo{author}{I.~Becker}, \bibinfo{author}{A.~Nill}, \bibinfo{author}{C.~J. Shultz}, \bibinfo{author}{J.~W. Gentry},
\newblock \bibinfo{title}{Male breadwinner ideology and the inclination to establish market relationships: Model development using data from germany and a mixed-methods research strategy},
\newblock \bibinfo{journal}{Journal of Macromarketing} \bibinfo{volume}{36} (\bibinfo{year}{2016}) \bibinfo{pages}{149--167}.
%Type = Article
\bibitem[{Johansson et~al.(2025)Johansson, Tykhonov, Alexandersson, Ferguson, Hanlon, Scharnhorst, and Osborne}]{johansson2025knowledge}
\bibinfo{author}{M.~Johansson}, \bibinfo{author}{V.~Tykhonov}, \bibinfo{author}{S.~Alexandersson}, \bibinfo{author}{K.~Ferguson}, \bibinfo{author}{J.~Hanlon}, \bibinfo{author}{A.~Scharnhorst}, \bibinfo{author}{N.~Osborne},
\newblock \bibinfo{title}{A knowledge base for arts and inclusion--the dataverse data archival platform as a knowledge base management system enabling multimodal accessibility},
\newblock \bibinfo{journal}{arXiv preprint arXiv:2504.05976}  (\bibinfo{year}{2025}).
%Type = Inproceedings
\bibitem[{Zeng and Mayr(2021)}]{zeng2021knowledge}
\bibinfo{author}{M.~L. Zeng}, \bibinfo{author}{P.~Mayr},
\newblock \bibinfo{title}{Knowledge organization systems (kos) in the semantic web. a multi-dimensional review},
\newblock in: \bibinfo{editor}{R.~P. Smiraglia}, \bibinfo{editor}{A.~Scharnhorst} (Eds.), \bibinfo{booktitle}{Linking Knowledge: Linked open data for knowledge organization and visualization}, \bibinfo{organization}{Ergon-Verlag}, \bibinfo{publisher}{Ergon}, \bibinfo{year}{2021}, pp. \bibinfo{pages}{34--63}. \DOIprefix\doi{10.5771/9783956506611-34}.
%Type = Article
\bibitem[{{Sally Chambers and Frédéric Lemmers}(2021)}]{8709832}
\bibinfo{author}{{Sally Chambers and Frédéric Lemmers}},
\newblock \bibinfo{title}{{Collections as Data}},
\newblock \bibinfo{journal}{{META (ANTWERPEN)}}  (\bibinfo{year}{{2021}}) \bibinfo{pages}{{36--37}}. \URLprefix \url{{https://www.vvbad.be/meta/meta-nummer-20213/collections-data}}.
%Type = Misc
\bibitem[{Jones et~al.(2023)Jones, Leggott, Lopez~Albacete, Pascu, Payne, Schouppe, Treloar, and IG}]{jones_2023}
\bibinfo{author}{S.~Jones}, \bibinfo{author}{M.~Leggott}, \bibinfo{author}{J.~Lopez~Albacete}, \bibinfo{author}{C.~Pascu}, \bibinfo{author}{K.~Payne}, \bibinfo{author}{M.~Schouppe}, \bibinfo{author}{A.~Treloar}, \bibinfo{author}{G.~O. R.~C. IG}, \bibinfo{title}{Gorc ig: Typology and definitions}, \bibinfo{year}{2023}. \DOIprefix\doi{10.15497/RDA00087}.

\end{thebibliography}

\end{document}